# Quantum Mechanics on a Ring: Continuity versus Gauge Invariance

Dr. Arthur Davidson
ECE Department
Carnegie Mellon University,
Pittsburgh, PA 15213, USA
artdav@ece.cmu.edu

**Abstract:** Remarkably we find that for a ring with linear boundary conditions such that the eigenvector and its derivative are continuous, there does not seem to be a way for the well-known de Broglie relation to be gauge invariant. Certain nonlinear boundary conditions assure gauge invariance, and lead to eigenfunctions with a discontinuous but differentiable phase and a continuous spectrum. A discrete subset of this spectrum forms a Hilbert space, while another subset is excluded by the nonlinear boundaries. We conclude that discontinuous momentum eigenfunctions are tenable, and that it is possible that quantum mechanics can have nonlinear boundary conditions in some circumstances.

## 1.    INTRODUCTION

Quantum dynamics have been characterized by fully linear equations and boundary conditions. Weinberg [1] offered a hypothetical nonlinearity and used atomic clock data on a hyperfine transition in the ground state of beryllium to project an upper limit on nonlinearity of $10^{-21}$ of the binding energy per nucleon. Experiments on the same system by Bollinger [2] et al extended the upper limit down by five orders of magnitude. Gisin [3] used reasoning in quantum communication to show that Weinberg nonlinearities would violate causality. However there is no literature -- other than by the present author [4] -- considering the possibility that nonlinearities could affect quantum systems through their boundary conditions. Here we show for a quantum system configured as a one dimensional ring that linear boundary conditions fail to produce an always gauge invariant de Broglie [5] relation. There seems to be no way out of this dilemma except to use nonlinear boundary conditions.  We show that this sacrifice of continuity and linearity at the boundary preserves a conventional Hilbert space with robust gauge invariant de Broglie properties.

First we will review the theory showing linearity and gauge invariance of the momentum eigenvalue for a one dimensional quantum domain with boundaries at infinity. We'll write the momentum eigenvalue equation in a gauge dependent form that arises from the arbitrary total time derivative that may be added to any mechanical Lagrangian without changing the classical equations of motion.[6] We'll show that in the quantum system identical eigenvalues result regardless of the gauge choice in this linear system.

Next, we'll consider a finite one dimensional ring, with the same momentum eigenvalue equation. Now the wavefunction is finite around the ring, and we must describe the way it interacts with itself at the boundary. We will show that the usual linear boundary conditions based on continuity are equivalent to a gauge dependent de Broglie relationship between the momentum eigenvalue and the wavelength. Then we'll show that alternative nonlinear boundary conditions cleanly restore gauge invariance for the ring. The invariant system forms a continuous spectrum of eigenvalues. A subset of the eigenfunctions will be allowed by the nonlinear boundary conditions to form a regular Hilbert space and be superposable. This subset is equivalent to the Bloch function [7] well established for extended periodic systems including a real number jump discontinuity in the phase at the boundary. The energy bands associated with the Bloch function would support quantized transitions to explain the experimentally observed quantized behavior of rings [8]. Thus nonlinear boundary conditions for the one dimensional ring maintain the essentials of Hilbert space and the superposition principle for a subset of eigenfunctions, along with gauge invariance, Hermiticity of the momentum operator, and discrete momentum transitions. Although strict continuity of the state vector is not maintained, all observables remain continuous, such as the probability density.

Josephson junctions are strong candidates for constructing qubits [9] for quantum computers. But these devices can be treated as particles on a strict 1D ring [10], and thus may have nonlinear boundary conditions and some unsuperposable eigenstates. How to fold this into the interpretation of quantum computation experiments remains a question. In appendix A it is shown that the mathematics of the strictly 1D ring is easily extended to a 1D ring embedded in 3D with a magnetic field. In appendix B, an explicit momentum operator with nonlinear boundary conditions is presented and diagonalized.

## 2. THE INFINITE ONE DIMENSIONAL CASE

Consider one coordinate dimension $x$ whose interval is infinite. The momentum operator for a neutral particle can be written:

$$\hat{p} = -i\frac{\partial}{\partial x} - \frac{\partial \xi(x)}{\partial x}. \qquad (1)$$

Units have been chosen so that $\hbar = 1$. $\xi(x)$ is a differentiable real function of x which may be taken to be the arbitrary function added to the classical Lagrangian[6]. For now, we assume $\xi(x)$ is continuous, but when we consider the ring geometry below this will be reconsidered. If we take the total time derivative of $\xi(x)$ we get $v(\partial \xi(x)/\partial x)$ where $v$ is the classical velocity of the particle. Since the canonical momentum is defined as the partial derivative of the Lagrangian with respect to velocity, we get (1) as the quantum mechanical momentum operator whose eigenvalues will correspond to the classical kinetic momentum $mv$. Also (1) satisfies the standard canonical commutation relation when paired with the operator for position, $\hat{x} = x$.

The eigenfunction complementary to Eq. 1 can be written

$$\psi(x) = \Gamma e^{i(nx+\xi(x))} \qquad (2)$$

where the number $n$ and the normalization constant $\Gamma$ are real.

As is well known in quantum mechanics, an operator applied to one of its eigenfunctions should yield a real constant eigenvalue multiplying the same eigenfunction. Thus:

$$\left(-i\frac{\partial}{\partial x} - \frac{\partial \xi}{\partial x}\right)\left(\Gamma e^{i(nx+\xi)}\right) = n\left(\Gamma e^{i(nx+\xi)}\right) \qquad (3)$$

so that $n$ by itself is the eigenvalue. That is, the derivative of $\xi$ cancels out of the eigenvalue expression. Even though Eq. 3 is written for a neutral particle, a specific functional choice for $\xi$ may be said to be part of the gauge for the momentum [6, 11] and $n$ is the gauge invariant eigenvalue. The boundary conditions play no role in this one dimensional problem extending far from the origin. The eigenfunctions extend over the whole interval, and there is a continuous spectrum of real eigenvalues. Since Eq. 3 is linear and homogeneous in the eigenfunction, and there are no effective boundary conditions other than normalization, this system is linear. All eigenfunctions are part of the Hilbert space.

Because of $\xi$ and $\partial \xi/\partial x$ this system has a momentum operator and eigenfunctions dependent on the choice of gauge. However, the eigenvalue equation returns a gauge invariant eigenvalue that is therefore measureable.

### 3. THE FINITE ONE DIMENSIONAL RING WITH LINEAR BOUNDARY CONDITIONS

Equations 1, 2, and 3 should apply to the one dimensional ring as well as the infinite line. But the eigenfunctions should now satisfy definite boundary conditions, namely the continuity and periodicity of $\phi(x)$ and $d\phi(x)/dx$ where $\phi(x)$ is a general wavefunction. These linear boundary conditions assure that any linear superposition of eigenfunctions also satisfies the boundary conditions.

If the x interval extends from $-\pi$ around the ring to $\pi$ then the eigenfunction phase in Eq. (2) must satisfy as the boundary condition

$$n + \frac{\xi(\pi,t) - \xi(-\pi,t)}{2\pi} = n + \frac{\Delta \xi}{2\pi} = m. \tag{4a}$$

where $m$ is an integer and $\Delta \xi$ is the jump discontinuity of $\xi(x)$ at the boundary. We re-write (4a) in the form of the de Broglie relation by noting that $n$ and $m$ behave as variables in (4a), not constants: $n$ is unambiguously the momentum eigenvalue; $m$ is both the winding number and reciprocal wavelength $2\pi/\Lambda$ for the eigenfunction. Using $\Lambda$ instead of $m$ we get

$$n \Lambda = 2\pi - \frac{\Delta \xi \, \Lambda}{2\pi}. \tag{4b}$$

If $\Delta \xi \equiv 0$ or equivalently if $\xi$ is always continuous then (4b) is the gauge invariant de Broglie relation for the ring in our units. If $\xi$ is not always continuous then (4b) is not gauge invariant. But the other term in the phase of the complex exponential in Eq. (2) is $nx$, which is always discontinuous for finite $n$. So we have to allow $\xi$ also to be discontinuous. Therefore we must conclude that continuous boundary conditions on a ring lead to a de Broglie relation that is not gauge invariant.

We now show why the function $nx$ must be differentiable even though it is discontinuous. Consider the phase in Eq. (2). If $n$ is finite the phase is discontinuous. If this means that the phase cannot be differentiated, then the only allowed eigenvalue is $n=0$. However, empirically a finite momentum is allowed, so $nx$ must be differentiable despite a discontinuity.

Actually a consistent derivative of a discontinuous phase is straightforward. It requires that the derivatives on either side of the discontinuity should approach the same value, and that there is a boundary condition at the point of interruption. In this case, a reasonable boundary condition is that the derivative should be continuous. But if the phase of the wavefunction can be differentiated despite a discontinuity, then it follows that the eigenfunction, which is the complex exponential of the phase, can also be differentiated despite a phase discontinuity of an arbitrary size. This means that there is no mathematical reason to require the continuity of the eigenfunction for either the infinite line in Section 2, or for the ring configuration.

We are forced to consider that linear continuous boundary conditions on a ring are not compatible with a gauge invariant de Broglie relation between momentum and wavelength, and that a discontinuous wave-function may still be differentiable.

## 4. THE FINITE ONE DIMENSIONAL RING WITH NONLINEAR BOUNDARY CONDITIONS

The obvious alternative boundary conditions to assure a gauge invariant de Broglie relation are continuity of the products $\varphi^*\varphi$ and $\varphi^* d\phi/dx$. Since these are products, they are nonlinear. We get the correct de Broglie relation because the phase itself plays no role. Note that if $\varphi(x) = \Gamma(x)e^{i\alpha(x)}$ then $\varphi^*\varphi_x = \Gamma\Gamma_x + i\Gamma^2\alpha_x$ where the subscript denotes partial differentiation. The imaginary part is the probability current density. Cleary, $\Gamma$, $\Gamma_x$, $\Gamma^2$ and $\alpha_x$ must be continuous and periodic. This condition implies that $\partial\xi/\partial x$ from Eq. 1 must be periodic. Linear boundary conditions allow $\alpha$ to have a $2\pi m$ jump discontinuity, while these nonlinear conditions allow any magnitude of real jump discontinuity in $\alpha$ at the boundary and yet have no discontinuities in measureables. We have already shown in Section 3 above that the eigenfunction must be differentiable even if the phase has a jump discontinuity.

The nonlinear boundary conditions will not change the Hermitian character of the momentum operator. The change from linear conditions can be accommodated by phase rotations of the momentum operator in opposite directions at the boundary in a way which preserves Hermiticity, but depends on which eigenfunction is operated upon. The character of the momentum operator as a matrix with nonlinear boundaries is developed in Appendix B. Figure B1 there shows computed momentum eigenvalues for the momentum operator with nonlinear boundary conditions.

The nonlinear boundary conditions are satisfied by the eigenfunction in Eq. 2 on the ring for all real *n*. Thus the eigenfunctions selected by the ring boundary have the same continuous spectrum as for the infinite line discussed in Section 2. However, with integration over the finite interval of the ring coordinate a subset of eigenfunctions within the continuous spectrum loses its orthonormality.

For the ring, not all eigenfunctions with real eigenvalues will form a Hilbert space. It is known, for example, that when two or more eigenfunctions with the form of Eq. 2, with different values of *n* are put in superposition, the resulting probability density will have periodic variation, and not all combinations of *n's* will permit matching the period to the interval of the ring. To make this explicit, choose one eigenfunction as the initial state with arbitrary real momentum eigenvalue *q*. Then we can go through all the other eigenfunctions, express their eigenvalues as *q* + *n*, and ask which subset of them will result in the periodicity of $\varphi^*\varphi$ and $\varphi^* d\phi/dx$ that matches the ring. The superposition will look like

$$\psi(x) = \Gamma e^{i\xi(x)} e^{iqx} \sum_{n=-\infty}^{n=+\infty} a_n e^{inx} \qquad (5)$$

where $\xi(x)$ is still the same gauge term in Eq. 1, and *n* is still the set of reals *a priori*. *q* is a constant selected from the real continuous spectrum of momentum eigenvalues, and the $a_n$ are a set of complex coefficients.

To apply our proposed boundary conditions to this sum in Eq. 5, we need to evaluate $\psi^* d\psi/dx$. It is straightforward to show that the (j k) term of this product will be

$$(jk) = i\Gamma^2 \left(n_j + q + \frac{\partial \xi}{\partial x}\right) a_{n_j} a_{n_k}^* e^{-i(n_k - n_j)x} \tag{6}$$

Here $n_j$ and $n_k$ are different possible eigenvalues in Eq. 5. It is clear that each term represented by Eq. 6 will be periodic over the ring if $\partial \xi / \partial x$ is periodic, and the $n_j$ and $n_k$ are restricted to integers.

The individual terms of Eq. 5, with $n$ restricted to integers form an orthonormal set and a conventional quantum Hilbert space. Since the measured kinetic energy of an eigenstate of this system is proportional to the square of *(q + n)*, with *q* continuous and n discrete, the Hilbert space will be that of a free particle in a system of quadratic energy bands, as discussed above in connection with Bloch's theorem. Evidently Bloch's theorem applies to a finite domain with the nonlinear boundaries, as well as to a periodic infinite domain [7].

The existence of energy bands proportional to *(q+n)$^2$* explains the experimental observation [8] of quantized behavior despite the continuous eigenvalue spectrum of the nonlinear momentum operator. Notice that for example, if q=1/2, then the energy eigenstates with n=0 and -1 are degenerate, allowing a discrete momentum change of $\hbar$ to occur. On the other hand if *q = 0* consistent with linear boundary conditions the favored momentum transition would be *2$\hbar$*, since the *n = +1* and *n = -1* states would be the lowest with degenerate energy levels.

It is interesting that this work predicts that mirror image rotation states of a ring such as eigenvectors with eigenvalues *(q + n)* and *– (q + n)* are not generally superposable. This is true even though each state is separately an eigenstate. The condition that these states are superposable is that q should be an integer multiple of ½. Some work [12] on qubits in superconducting loops may need reexamination.

## 5. COMPARISON TO THE WEINBERG NONLINEARITY

The well-known Weinberg nonlinearity [1] has been shown to conflict with causality [3] because that class of nonlinearity mixes eigenvectors and changes their directions in Hilbert space. The nonlinear boundary conditions discussed here will not have this property. Cross terms will be generated at the boundary from the $\varphi^* \varphi$ nonlinearity, but these only contribute to eigenfunction selection, and do not feed back into the eigenfunctions themselves. The dynamic variable remains the complex wavefunction evolving according to the still perfectly linear Schrödinger equation.

## 6. SUMMARY

In the one dimensional ring, gauge invariance of the de Broglie relation is surprisingly not deduced from the usual linear boundary conditions, but consistent with certain nonlinear ones. The ring with nonlinear boundaries where certain products of wave functions are continuous has a continuous infinity of non-orthonormal momentum eigenfunctions with a continuous spectrum of real gauge invariant eigenvalues. There is also an arbitrary real valued jump discontinuity in the phase of the wave-function at the boundary. A subset of these eigenfunctions is allowed by the nonlinear boundary conditions in orthonormal superposition equivalent to Bloch's theorem. This subset remains superposable such that a ray in Hilbert space, once set up, is unperturbed by the boundary nonlinearity. Systems describable as a particle on a ring will not have discrete energy levels, but energy bands instead. Transitions between these bands will account for the observed behavior of quantum rings.

There will be another subset of eigenfunctions that are unsuperposable. It may be that Schrodinger's cat is never found in superposition because the two states of the cat are not orthonormal due to nonlinear boundaries. Likewise, superposability cannot be assumed for all qubits of a quantum computer.

The consistency of nonlinear boundaries for this system suggests that nonlinearity may be important for the description of a quantum system coupled to its environment, possibly along the line of Fermi [13], Kostin [14,15] N. Gisin [16], Davidson and Santhanam [17], and others. Certainly the Josephson junctions used in some important quantum computer experiments [9, 12, 18, 19] need to be re-examined, since these devices and some circuits can be modeled as a particle on a ring [20] such as considered here.

We have shown evidence that the usual linear boundary conditions based on a continuous wave function are insufficient for quantum mechanics on a ring. We have suggested nonlinear boundary conditions that are sufficient. Whether they are necessary may require experiment. The acceptance of the nonlinear boundaries proposed here requires the acceptance of a discontinuous eigenfunction, which is a significant departure from conventional theory, but allowed both mathematically and physically.

## ACKNOWLEDGEMENTS

Dr. Alan Kadin of Princeton Junction New Jersey provided constructive comments as this work has progressed. Prof. James A. Bain of Carnegie Mellon University provided interest and encouragement. My wife Janis has supported long term my persistence in studying the tension between continuity and gauge invariance on a quantum ring.

# Appendix A: Gauge transformations on a 1D ring in 3D with electromagnetism.

The purpose of this section is to develop the theoretical treatment of a quantum 1D ring embedded in *3D* with full classical electromagnetism. The results will be shown to be compatible with treatment of an isolated 1D ring without electromagnetic interaction presented in the body of this work.

Here are the momentum operator and eigenfunction for a charged particle in 3D in an electromagnetic field represented by a vector potential in cylindrical coordinates.

$$\vec{P}_{op} = [-i\hbar\vec{\nabla} - \overrightarrow{eA} + \vec{\nabla}(e(1-\gamma)\chi - \xi)] \tag{A1}$$

$$\psi(r, \varphi, z) = \Gamma e^{\frac{i}{\hbar}(\vec{n}\cdot\vec{r} + e\gamma\chi + \xi)} \tag{A2}$$

The electron charge is e. $\xi$ and $\chi$ are scalar functions of the coordinates (r,ϕ, z.). $\vec{A} - \vec{\nabla}\chi$ is the vector potential, with information on both the electric and magnetic fields. $\vec{\nabla}\chi$ represents part of the vector potential whose curl is zero. $\xi(r,\varphi,z)$ in the operator and wavefunction is inserted arbitrarily consistent with the term that may be added to any Lagrangian, as discussed in the Introduction and Section 2 of the main body of this paper. At this point, $\xi$ is a 3D scalar in the 3D equations A1 and A2. The reason for its presence will become apparent when the wave function becomes 1D in the next paragraph. $\gamma$ is an arbitrary real number [0,1], which partitions $\chi$ between the operator and the wavefunction. When applying the operator to the eigenfunction, $\xi$ cancels out of the eigenvalue, while $\chi$ contributes to it. Moving part of the vector potential to the wave function is standard gauge theory. Byers and Yang [A1] in their theory paper on flux quantization moved part of their vector potential (also denoted by $\vec{\nabla}\chi$) to the wavefunction in their Eq. 4, and then used linear boundary conditions based on the new wave function phase. What is not clear in their paper is what boundary conditions to use in their original gauge.

Now restrict the wavefunction to a 1D ring at the edge of an appropriate surface orthogonal to $\hat{z}$. In this case $\psi$ and $\xi$ become 1D scalar functions, but the vector potential remains 3D. We have for the momentum eigenvalue equation on the 1D ring at radius r:

$$\vec{P}_{op}\psi(\varphi) = \left(n - e\left(A(\varphi) - \frac{1}{r}\frac{\partial}{\partial\varphi}\chi\right)\right)\Gamma e^{\frac{i}{\hbar}(n\varphi + e\gamma\chi + \xi)} \tag{A3}$$

In Eq. A3, we seek solutions where the factor multiplying the wave function is a real constant eigenvalue $\lambda = n - e(A(\varphi) - (1/r)(\partial\chi/\partial\varphi))$. If we take the contour integral of λ over the 1D ring at radius r and solve for $\lambda$, we get $\lambda = (n + \frac{e\Phi_T}{2\pi r})$ where $\Phi_T$ is the magnetic flux through the 1D ring. Since A and ∇χ are both part of the 3D vector potential defined over a surface they contribute to Stokes theorem. In the 1D ring geometry *ξ(φ)* plays no role in the magnetic field for two reasons: first because it cancels out of the momentum eigenvalue, and second because it is defined only on the ring, not over the surface enclosed, thus nullifying Stoke's theorem for $\xi$ in the eigenfunction phase.

Eq. A3 is in MKS units; Eq. 3 is dimensionless. Nonetheless they may be readily compared. Clearly, the magnetic field contributes to the eigenvalue in Eq. A3. The terms involving ξ clearly play the same non-magnetic role in each equation. The presence of an arbitrary fraction γ in the exponent in Eq. A3 reinforces the idea that the wave function phase should not be involved in boundary conditions for gauge invariant eigenvalues.

A final observation: a neutral particle on a ring embedded in 3D can be represented by the same 3 equations, A1, A2 and A3, with e$A$ and e$\chi$ set to zero. ξ, however, should remain as presented in equations 1, 2, and 3 justified by Lagrangian mathematical physics.

**Appendix A REFERENCE**

[A1] N. Byers, and C.N. Yang, *Phys. Rev. Lett.* 7, 2, 46 (1961).

# Appendix B: Construction of the momentum operator with nonlinear boundary conditions

In Appendix B we develop the alternative nonlinear boundary conditions for the momentum operator matrix for a ring domain. We show that this matrix with nonlinear boundary conditions works in the usual linear static eigenvalue equation for momentum.

We first discuss the well known matrix form of the momentum operator in the spatial representation for linear periodic boundary conditions, $L_{jk}^{lin}$. This matrix is of rank $r$, while $j$ and $k$ are the row and column indices. The matrix elements will be chosen to correspond in the limit of small coordinate differences to the differential momentum operator used in Eq. 1 with $\partial \xi/dx=0$. $L_{jk}^{lin}$ will be Hermitian and linear, with linear homogeneous boundary conditions.

Next, $L_{jk}^{lin}$ will be morphed into $L_{jk}^{nl}$, the matrix with nonlinear boundary conditions. Then, a commercial matrix solver will be used to find the eigenvalues of $L_{jk}^{nl}$ for $r = 20$. The predicted continuous and discrete parts of the eigenvalue spectrum appear, confirming consistency in using nonlinear boundary conditions in the ordinary eigenvalue equation.

**The linear periodic momentum operator:** The matrix elements of $L_{jk}^{lin}$ may be specified as follows: they are zero except for the super-diagonal, the sub-diagonal, and the two cells at the extremes of the minor diagonal. If we take $\partial \xi/dx=0$, then the diagonal is zero.

The non-zero elements are: Super-diagonal elements are $L_{j,j+1}^{lin} = -\frac{i}{2dx}$; Sub-diagonal elements are $L_{j,j-1}^{lin} = +\frac{i}{2dx}$. The minor diagonal ends are;

$$L_{1,r}^{lin} = +i/(2dx), \quad L_{r,1}^{lin} = -i/(2dx) \tag{B1}$$

The increment represented by one segment of the ring is $dx=2\pi/r$. The central difference method is used to approximate the derivative, preserving Hermiticity.

It is simple to prove that as $dx \to 0$ this operator applied to a column vector will approximate the first derivative with respect to x, multiplied by (-i) as required by standard quantum mechanics, for units chosen to make ℏ = 1. Notice the values in the (1, r) and (r, 1) corners give the operator linear periodic boundary conditions. The operator is linear and homogeneous everywhere, including the boundary.

**Morphing to nonlinear boundary conditions:** If as discussed in section 4 the nonlinear boundary conditions are that $\Gamma$, $\Gamma_x$, and $\alpha_x$ are smooth and periodic, while $\alpha$ has an arbitrary jump discontinuity, then we need a way to differentiate the wavefunction across the boundary where α jumps. Since $L_{jk}^{lin}$ differentiates across the boundary twice, once in its first row, and once in its last, the salient action for $L_{j,k}^{nl}$ at the boundary is to rotate the (1,r) and (r,1) cells in opposite directions by an angle $\Delta\alpha$:

$$L_{1,r}^{nl} = +ie^{i\Delta\alpha}/(2dx) \quad \text{and} \quad L_{r,1}^{nl} = -ie^{-i\Delta\alpha}/(2dx) \tag{B2}$$

$\Delta\alpha$ is the discontinuity of the phase $\alpha$ at the boundary. So if the vector operated on has $\alpha = qx$ Then $\Delta\alpha = 2\pi q$ where $q$ is the same as in Eq. 5.

To summarize, $L_{jk}^{lin}$ and $L_{jk}^{nl}$ have identical diagonals, super diagonals, and sub diagonals. The difference is that the ends of the minor diagonal in $L_{jk}^{nl}$ are rotated by an amount $\Delta\alpha$ that depends on the jump discontinuity in the phase of the state vector being operated on.

$L_{j,k}^{nl}$ remains homogeneous in the state vector, but nonlinear, since two of its elements depend on the jump discontinuity in the phase of the state vector. The x derivative of the state vector will be continuous everywhere except at the jump discontinuity, which is what is needed for gauge invariance. It is easy to prove that different eigenfunctions of $L_{j,k}^{nl}$ will be orthogonal if they share the same discontinuous jump in phase.

**Eigenvalue solutions:** We used the "eig" function in Matlab 2008a (The MathWorks, Natick, MA) to diagonalize $L_{j,k}^{nl}$ with r= 20 for eleven values of $\Delta\alpha$ between zero and p. Figure B1 plots the numerical value of the "eig" solutions for the eigenvalues vertically against $\Delta\alpha$ on the horizontal axis. The "eig" function also produced some spurious eigenvalues, which were discarded. The lines in Fig. B1 are what is expected theoretically from the nonlinear boundary conditions. The circles are the result of the "eig" diagonalization.

If λ is the real eigenvalue of the operator, the solution for the nonlinear boundaries is $\lambda = ((\Delta\alpha/2\pi) + n)$ where n is an integer. This is essentially what emerged from the Matlab diagonalization. When $\Delta\alpha \neq 0$ the eigenvalues become momentum bands. An arbitrary state will have a single real value of $\Delta\alpha$, but multiple n states in the superposition. In contrast, linear boundary conditions allow $\Delta\alpha = 2\pi m$ only, while ignoring gauge issues.

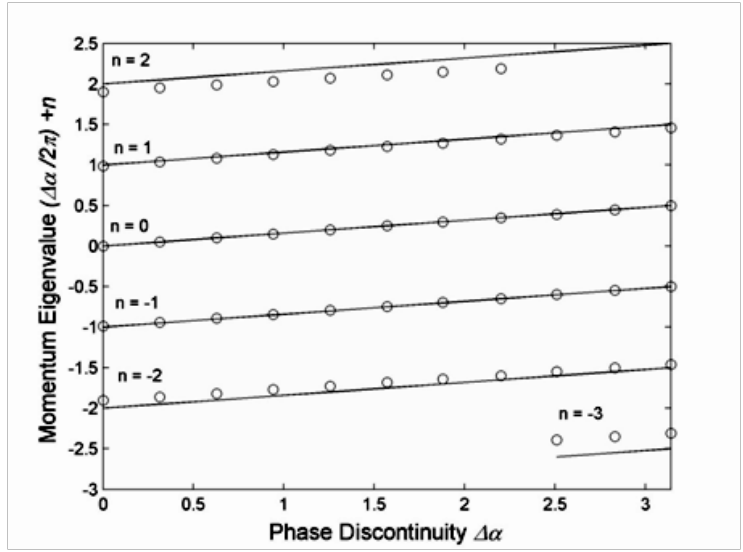

**Figure B1. Momentum eigenvalues plotted against the state vector phase jump discontinuity for the momentum operator with nonlinear boundary conditions.**